\documentclass[a4paper,11pt]{article}
\usepackage{jinstpub} % for details on the use of the package, please see the JINST-author-manual
\usepackage{lineno}
\usepackage{siunitx}
\title{\boldmath PALM - Precision Attenuation Length Measurement in Liquid Scintillators}

\author[a,1]{Vincent Rompel\note{Corresponding author.},}
\author[a]{Sabrina Franke,}
\author[a]{Florian Kübelbäck,}
\author[a]{Lothar Oberauer,} 
\author[a]{\mbox{Luca Schweizer,}} 
\author[a]{Korbinian Stangler,}
\author[a]{Hans Th. J. Steiger,}
\author[a]{Matthias Raphael Stock}
\author[a]{and Andreas Ulrich}

\affiliation[a]{Technical University of Munich, TUM School of Natural Sciences,\\
Physics Department, James-Franck-Str. 1, 85748 Garching, Germany}

\emailAdd{vincent.rompel@tum.de}

\abstract{
Future neutrino experiments at low energies such as~JUNO or~\textsc{Theia} will use large volume homogeneous liquid scintillator detectors. The optical attenuation length of the liquid is of uttermost importance for the successful realization of these experiments. At TU Munich a new optical spectrometer (Precision Attenuation Length Measurement (PALM)) has been set up in order to measure the light attenuation of liquids which can be used for these types of experiments up to around \SI{100}{\m} in the wavelength region between \SI{400}{\nm} and \SI{1000}{\nm}. The setup features an optical imaging system with a long focal length, which allows for in-situ monitoring of the beam stability during the measurement. The setup capability as well as the reproducibility of its results has been demonstrated at two different wavelengths of \SI{430}{\nm} and \SI{500}{\nm}.}

\begin{document}
\maketitle
\flushbottom

\section{Introduction}
\label{sec:intro}

Homogeneous, non-segmented large volume liquid scintillator detectors exhibit essential advantages for neutrino experiments at low energies in the sub-GeV regime.
The main advantage is the self-shielding capability of the large volume detector against external gamma and neutron radiation.
The liquid scintillator, as the most inner part of the whole detector, can be produced with extremely low concentrations of radioactive elements as was demonstrated e.g. in the solar neutrino experiment Borexino~\cite{Bx-pur}.
In addition, it acts as an active shielding against external background radiation, and by introducing a so-called fiducial volume in the data analysis a very clean inner zone will be defined which is then used to search for rare neutrino interactions.

Currently, the upcoming~JUNO~\cite{Juno} experiment in China using $20\,\text{kton}$ of a liquid scintillator is the most prominent example of these types of detectors, but there are projects such as~\textsc{Theia}~\cite{Theia} in the US, where even larger volumes are discussed.
In all this detectors scintillation light has to overcome long distances on the order of tens of meters in order to reach the photo-sensors, which are kept outside the scintillation zone.
Therefore, the light attenuation length $\Lambda$ of the liquid has to be also in the same range.
Relevant is the value of $\Lambda$ with wavelengths 
$\lambda \sim $ \SI{430}{\nm}, 
as liquid scintillators usually emit light in the near UV-region.
The value of $\Lambda$ at these wavelengths is one of the key parameters in designing large-volume scintillation detectors.

In this paper we report about a new set-up at TU Munich named PALM ('Precision Attenuation Length Measurement') which is able to measure attenuation lengths in the relevant range with high accuracy.
Furthermore, PALM is capable of measuring the attenuation length of a liquid in a wide range of wavelengths.
This can be important 
as light attenuation occurs due to absorption or scattering and the underlying effect cannot be identified in a single measurement.
However, one can hope to do so by analyzing the attenuation length of a sample as a function of wavelength, 
as the cross-section of some processes like Raleigh scattering obeys a very distinct wavelength dependency.

\section{Experimental Setup}

\begin{figure}[htbp]
    \centering
    \includegraphics[width=\textwidth]{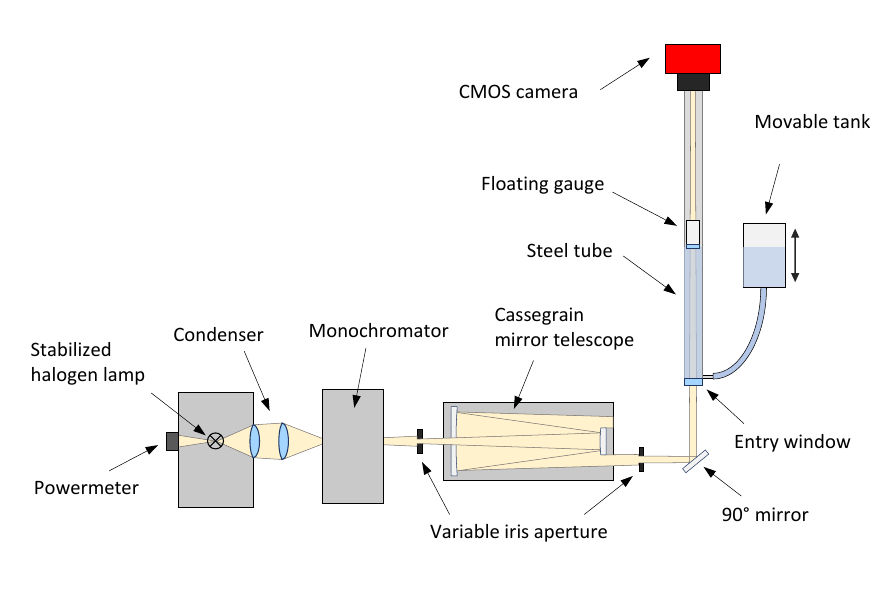}
    \caption{Schematic representation of the PALM experimental setup (not to scale). Details in the text.} 
    \label{fig:PALM_schema}
\end{figure}

\begin{figure}[htbp]
    \centering
    \includegraphics[width=0.9\textwidth]{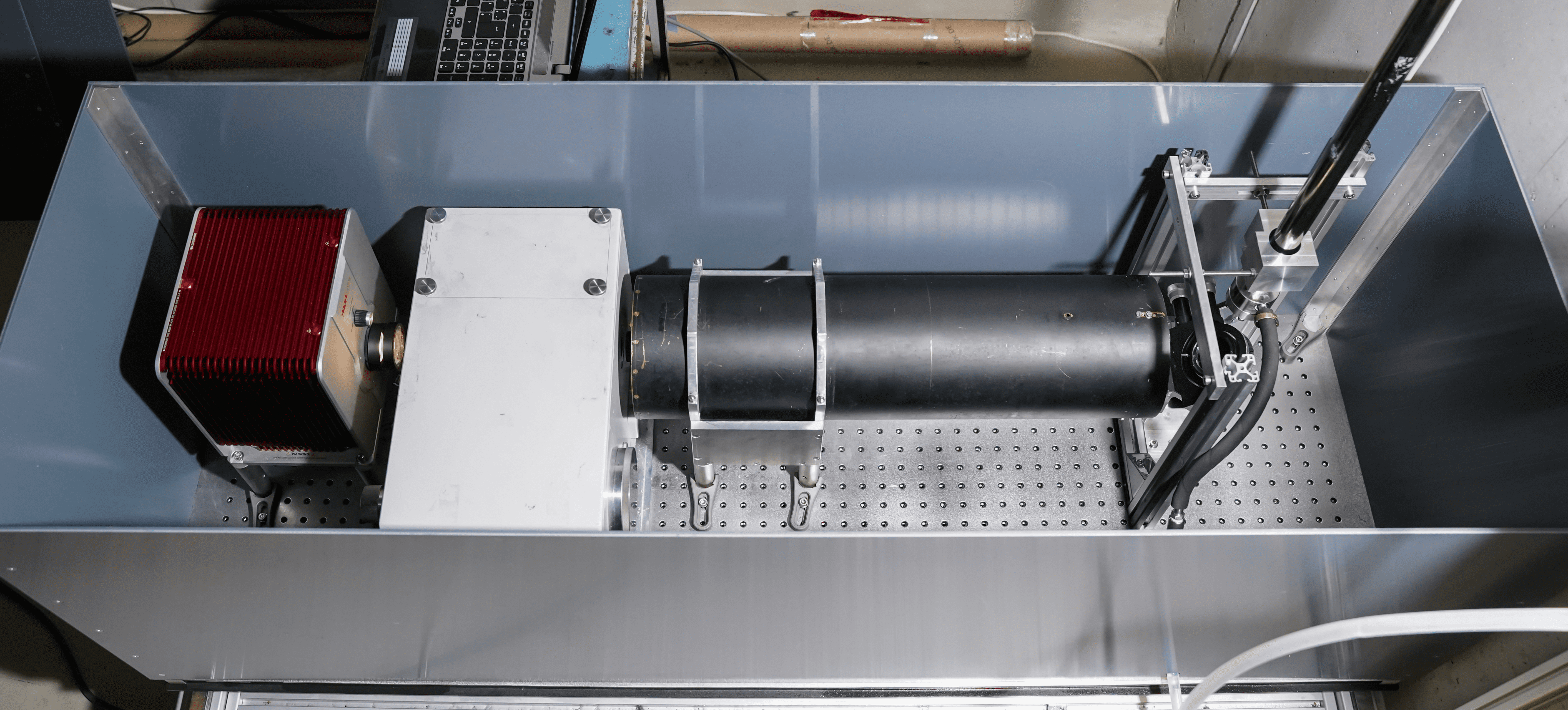}
    \caption{Top view of the PALM dark box. The halogen lamp, monochromator, telescope, iris diaphragm, and the \SI{90}{\degree} mirror with the bottom part of the tube can be seen.}
    \label{fig:Top_PALM}
\end{figure}

The attenuation length $\Lambda$ of a material is defined as the distance $x$ of a light beam, after which its intensity decreases to $1/e$ of the incident intensity $I_0$, through scattering and absorption. This can be described by the Beer-Lambert-Law:
\begin{equation}
    I(x) = I_0 e^{-x/ \Lambda}  
    \label{equ:Lambert}
\end{equation}
Therefore, the attenuation length of a medium can be determined by measuring the residual intensity of a light beam after it has traveled various distances in this medium. In general, the attenuation length is strongly wavelength dependent.
Figure \ref{fig:PALM_schema} shows a schematic drawing of the PALM setup. The light from the halogen lamp is coupled into the monochromator via a quartz glass condenser. The reference lamp intensity is monitored by a power meter attached to the back of the housing. The monochromatic light passes into a Cassegrain mirror telescope, which focuses the beam. Next, it passes through an iris diaphragm, a finely adjustable~\SI{90}{\degree} mirror, and a UV-transparent glass window into the \SI{3}{\m} long stainless steel tube. The tube is connected via PTFE tubing to the sample tank. Its height can be adjusted to alter the liquid level in the tube and, consequently, the distance the light travels through the sample. This allows to measure the light intensity on top of the tube, which is done with a CMOS camera. A floating gauge can be placed inside the tube on top of the liquid to prevent the formation of waves. The setup is partially placed inside a darkbox to block ambient light. \mbox{Figure \ref{fig:Top_PALM}} shows the top view of the darkbox.

%\begin{figure}[htbp]
%    \centering
%    \includegraphics[width=0.4\textwidth]{Pictures/experimental_setup_01.png}
%    \caption{Front view of the PALM experiment. The camera is attached to the holding structure on the top. The height scale is located on the right of the sample tank, parallel to the tube.}
%    \label{fig:Front_PALM}    
%\end{figure}

\subsection*{Optics}

The setup consists of the following main optical components, which are all UV light compatible:
\\ The Thorlabs SLS301 Stabilized Tungsten-Halogen Light Source is used. This lamp provides a highly stable power output with fluctuations of only \SI{0.05}{\percent} after an initial 45-minute warm-up phase. It is mounted in a protective casing that gradually warms up over time, ensuring that the power output remains stable despite fluctuations in the surrounding temperature. At the front of the casing, a quartz glass condenser is attached to collimate the light emitted by the light source. The light bulb has a typical lifespan of 1000 hours.
\\ The Thorlabs PM16-121 power meter is employed to monitor the lamp intensity. It has a resolution of $<\SI{10}{\nW}$ within a range from $<\SI{50}{\nW}$ up to $<\SI{500}{\mW}$, which allows for the measurement and subsequent correction of any remaining fluctuations in the intensity of the light source. It is mounted directly at the back of the lamp housing.
\\ The Ocean Insight Maya2000 Pro spectrometer is used to check the output light spectrum. The spectral range of the spectrometer reaches from \SI{165}{\nm} to \SI{1100}{\nm}. The Zeiss MM3 double-prism monochromator is used to select a specific wavelength. It is capable of providing monochromatic light between \SI{180}{\nm} and \SI{3500}{\nm} and has an adjustable slit width up to \SI{2}{\mm}. A quartz glass focusing lens positioned after the telescope focuses a portion of the incident monochromatic light into the spectrometer.
\\ The Atik Apx60 CMOS camera is employed to measure the remaining light intensity. It features a monochromatic full-frame $24\times36\,\text{mm}$, no-amp glow sensor, and an electronic shutter, which reduces fluctuations in the exposure time compared to mechanical shutters. The sensor can be cooled to~\SI{-20}{\degree}C (at~\SI{20}{\degree}C ambient temperature), reducing dark noise, which, in conjunction with low read-out noise, allows precise intensity measurements to be performed. Furthermore, the quantum efficiency is specified within the range between \SI{400}{\nm} and \SI{1000}{\nm}. Hence, the camera can be used for our purposes as liquid scintillators have a maximum of light emission at about \SI{430}{nm}. Using a camera as a light detector enables the beam position and optical quality to be monitored during measurement, which ensures measurement stability.

\subsection*{Mechanics}

\begin{figure}[htbp]
    \centering
    \begin{minipage}[b]{0.5\linewidth}
        \centering
        \includegraphics[width=0.87\linewidth]{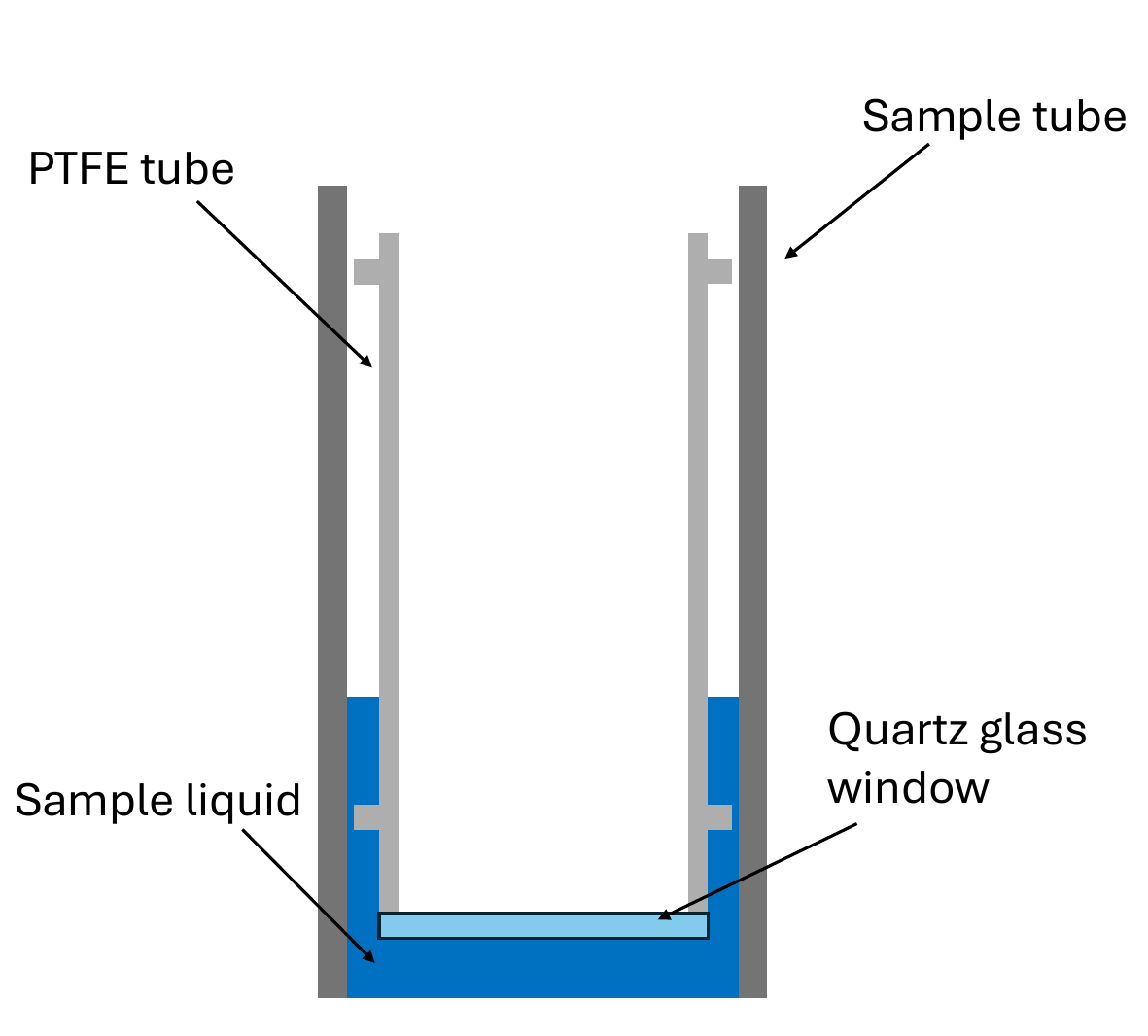}
        \caption{Scheme of the working principle of the floating gauge. It is placed on top of the sample inside the tube and than immerses to a certain depth into the sample. The rings on its outside guide the device along the tube wall and stabilize its position. The light beam passes through the window at the bottom.}
        \label{fig:schwimmer_workingprinciple}
   \end{minipage}
   \hspace{0.1\linewidth}
   \begin{minipage}[b]{0.3\linewidth}
        \centering
        \includegraphics[width=0.87\linewidth]{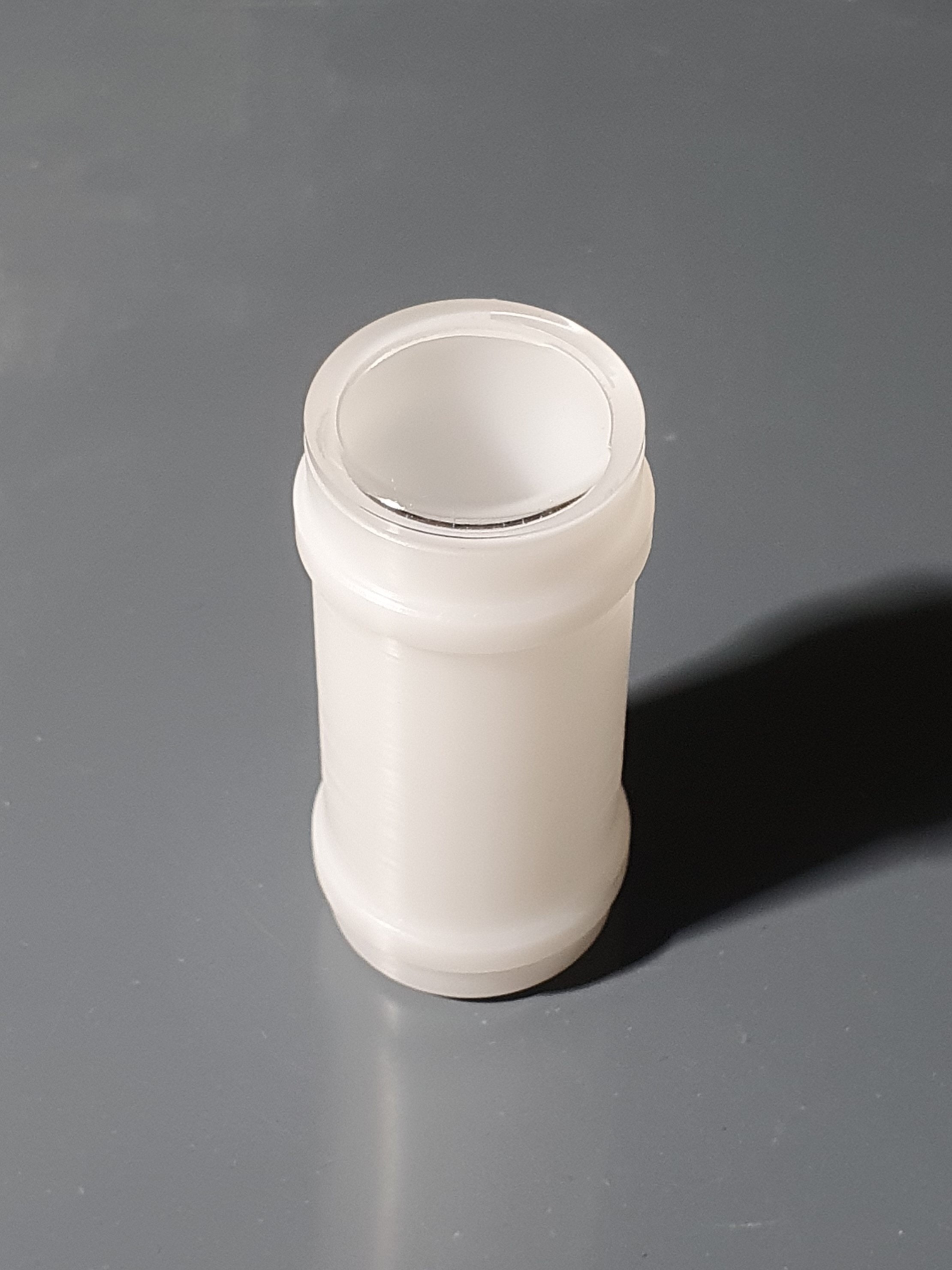}
        \caption{Picture of the floating gauge. It is made of PTFE for its density and chemical properties. A quartz glass window is glued on one end of the tube with epoxy glue.}
        \label{fig:schwimmer_picture}
    \end{minipage}
\end{figure}

To decouple the system from external vibrations, the base plate is placed on tennis balls, and the tube is hold by vibration-absorbing brackets. PTFE tubing is used to connect the tube to the tank, which consist of milled aluminum top and bottom plates and an acrylic glass wall.
\\ The sample tube is made from passivated stainless steel with an anti-reflective coating on the inside. The former prevents any reactions or contamination with the sample, whereas the latter reduces the amount of scattering light to a negligible level. This is important as scattering light is not considered in the data analysis. The diameter of approximately \SI{15}{\mm} allows for a small sample volume of approximately \SI{2}{\L}. PTFE tubing is used to connect the sample tube to the tank.
\\ A floating gauge is used to mitigate the impact of vibrations on the measurements. It sits on the surface of the sample and exerts light pressure to flatten it. It is constructed from a hollow \SI{20}{\cm} long PTFE cylinder. Protruding rings on its outside, with a slightly smaller radius than the sample tube, guide it and stabilize its position in the tube. A quartz glass window is glued to the bottom with epoxy glue for the light to pass through. Figure \ref{fig:schwimmer_workingprinciple} illustrates the floating gauge's working principle and Figure \ref{fig:schwimmer_picture} shows a picture.

\subsection*{Data Evaluation}

The total intensity of the light beam is obtained by summing the pixel intensities $I_{ij}$ of the whole sensor, where $i$ and $j$ denote the line and column of the pixel. To minimize statistical fluctuations, multiple images were captured per measurement point and averaged. Immediately after the measurement is completed, the background and dark noise are determined by taking multiple pictures with the light source turned off. The averaged background intensity $\langle I \rangle_{ij}^{\text{dark}}$ is then subtracted pixel by pixel. The reduced pixel intensity $I_{ij}^{\text{r}}$ given by
\begin{equation}
    I_{ij}^{\text{r}} =  \Big[\langle I \rangle_{ij}-\langle I \rangle_{ij}^{\text{dark}} \Big] \, \text{.}
\end{equation}
For light source fluctuations, this reduced intensity is adjusted by a factor $\omega$, based on the power meter readings. The calibrated pixel intensity $I_{ij}^c$ is given by
\begin{equation}
   I_{ij}^{\text{c}} =  I_{ij}^{\text{r}} \cdot \omega \quad \text{with} \quad \omega = \frac{\langle I_{\text{pm}}\rangle}{I_{\text{pm}}} \, \text{,}
\end{equation}
where $I_{\text{pm}}$ is the intensity of the lamp determined by the power meter during measurement of the individual data point and $\langle I_{\text{pm}}\rangle$ is the mean intensity of the light source during the whole measurement duration.

\section{Camera Calibration}

To ensure the measurement stability, the camera is calibrated and tested. The linearity of the sensor is checked, and the gain values of the individual pixels are determined. The calibration setup is realized in a non-reflective dark box and uses a stabilized halogen lamp on an optical bench. The distance $d$ between the lamp and the camera can be varied. Then, the intensity of each pixel $I_{ij}$, which is given after the dark noise correction, is
\begin{equation}
    I_{ij} = \frac{I_{0,ij}}{d^2} \, \text{,}
    \label{equ:Intensity_calib}
\end{equation}
with the normalized intensity $I_{0,ij}$. The gain of each pixel is then calculated by comparing the fit values of $I_0$ of the whole sensor with the values $I_{0,ij}$ of the individual pixel fits. The gain for each pixel are then given by $g_{ij} = I_{0,ij}/{I_{0}}$.

%\\ Figure \ref{fig:gainfactor_sensor} depicts the position of the pixels with their respective gain factor on the sensor of the camera. The gain factors follow a pattern on the sensor. In the center and the edges of the sensor, regions with lower gain factors can be seen. Additionally, some small dots with a higher gain factor can also be seen. These dots stem from dust particles on the sensor during the calibration measurement. 

\begin{figure}[htbp]
    \centering
    \includegraphics[width=0.8\textwidth]{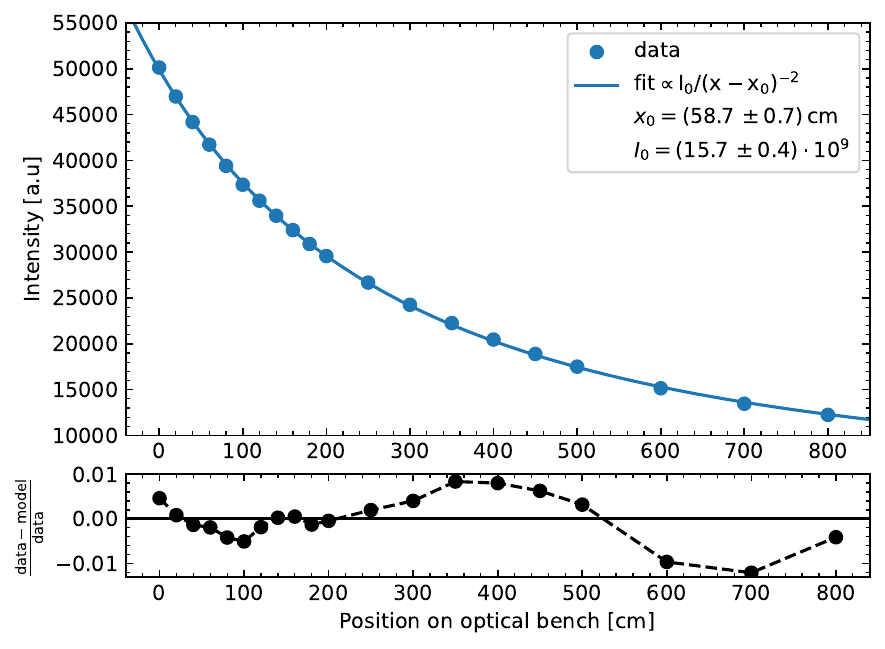}
    \caption{The mean pixel intensity of the CMOS sensor as a function of the position of the light bulb on the optical bench. Additionally the fit function $I \propto I_0/(x+x_0)^2$ is shown. Here, $x_0$ is the distance from the sensor to the optical bench, $x$ is the distance of the lamp on the bench, and $I_0$ is a normalization parameter. The normalized residuals are plotted in black.}
    \label{fig:linearity_pixel}
\end{figure}

\begin{figure}[htbp]
    \centering
    \includegraphics[width=0.8\textwidth]{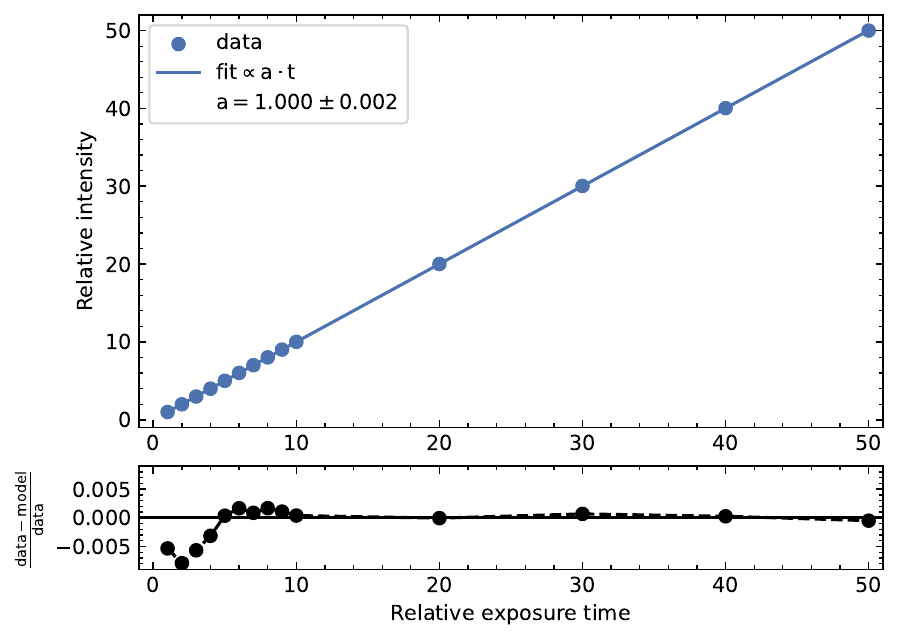}
    \caption{The relative intensity, as measured by the camera, as a function of the relative exposure time. Furthermore, the fit function $I \propto t$, where $t$ is the exposure time, is plotted. The normalized residuals are shown in black.}
    \label{fig:linearity_exposure}
\end{figure}

The standard deviation of the resulting distribution of gain factors is $\sigma = \SI{1,8}{\percent}$. Since the gain of the individual pixels is very similar and the beam covers many pixels during the attenuation length measurement, the influence of the difference in gain factors can be neglected.
\\ The linearity of the sensor is crucial for a precise measurement of the attenuation length, therefore two independent methods are used to ensure this. In the first approach, the data derived from the gain calibration are used. The global pixel fit is analyzed to determine if there is any deviation from the expected function. In Figure \ref{fig:linearity_pixel} the mean pixel intensity is plotted as a function of the lamp position, together with the fit according to equation \ref{equ:Intensity_calib} and the residuals.
\\ In the second method, the lamp is positioned in a fixed position at the front of the optical bench, and the exposure time of the camera is increased in steps from \SI{0.01}{\s} to \SI{0.5}{\s}. It is checked whether deviations from the expected linear increase in intensity with the
exposure time of the camera exist. Figure \ref{fig:linearity_exposure} shows the relative intensity plotted against the relative exposure time, together with a linear fit function and the corresponding residuals.
The fitted lines for both methods match the data very well and the normalized residuals show minimal deviation from the fit. Consequently, it can be concluded that there is no evidence of a deviation from linear sensor behavior within these calibration measurements.

\section{Experimental Results}
\label{sec:results}

\begin{figure}[htbp]
    \centering
    \includegraphics[width=0.8\textwidth]{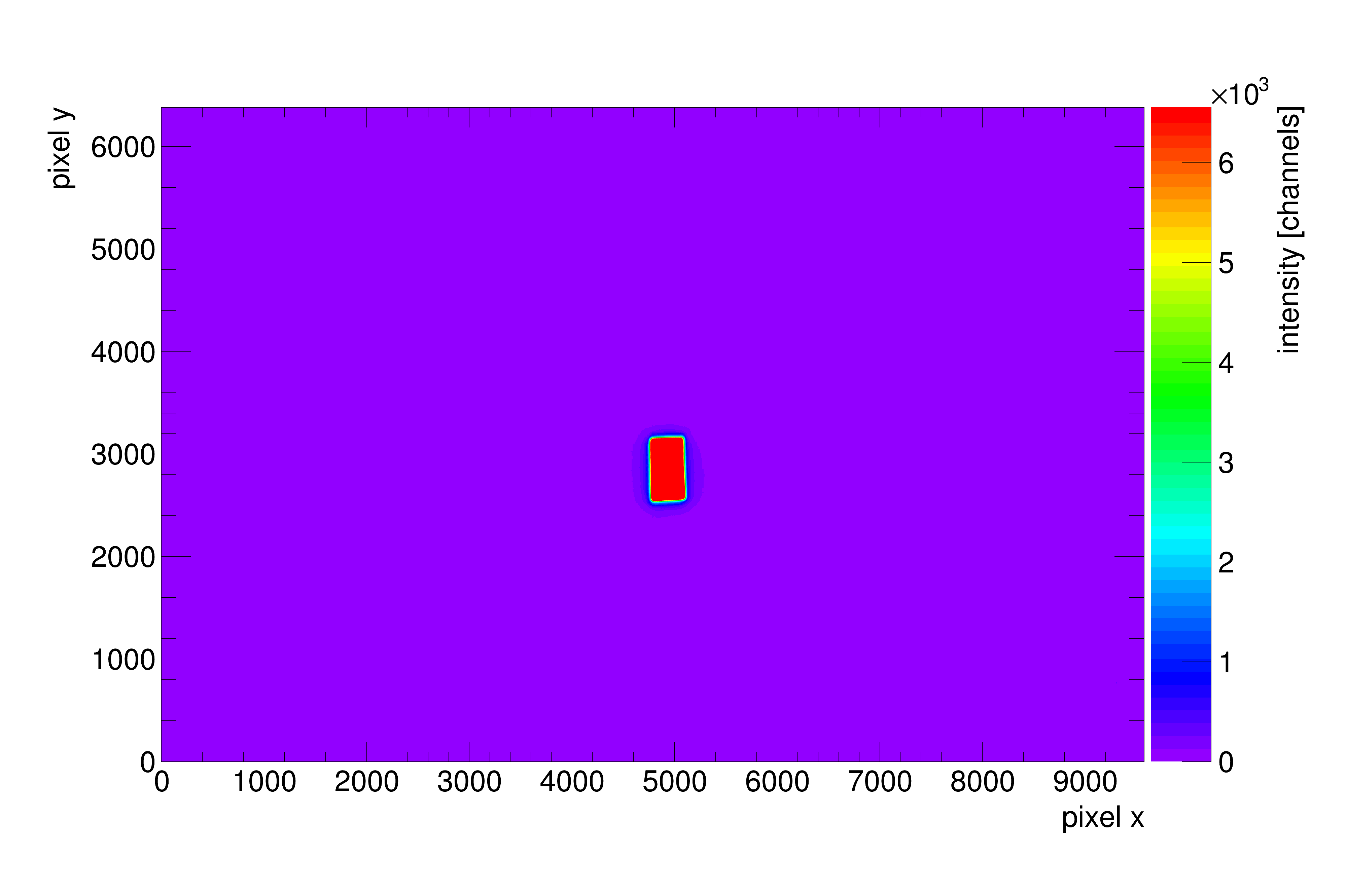}
    \caption{Focused and stable beam during the measurement}
    \label{fig:beam}
\end{figure}

All measurements were performed with the camera sensor cooled to \SI{-15}{\degree}C. In order to prepare a measurement, the lamp was thermalized and stabilized, the tank was brought to the maximum fill height, and the floating gauge was placed on top of the sample. The overall stability of the measurement was monitored in-situ by observing the optical images of the beam as demonstrated in \mbox{Figure \ref{fig:beam}}. The shape of the beam represents the slit of the monochromator. For a wavelengths of \SI{430}{\nm} and a slid width of \SI{2}{\mm} the first diffraction minimum occurs at $\sim \SI{0.65}{\mm}$, therefore diffraction effects can be neglected. The beam diameter is consistently less than \SI{5}{\mm}, making it significantly smaller than both the sensor and the tube.

\begin{figure}[htbp]
    \centering
    \includegraphics[width=0.8\textwidth]{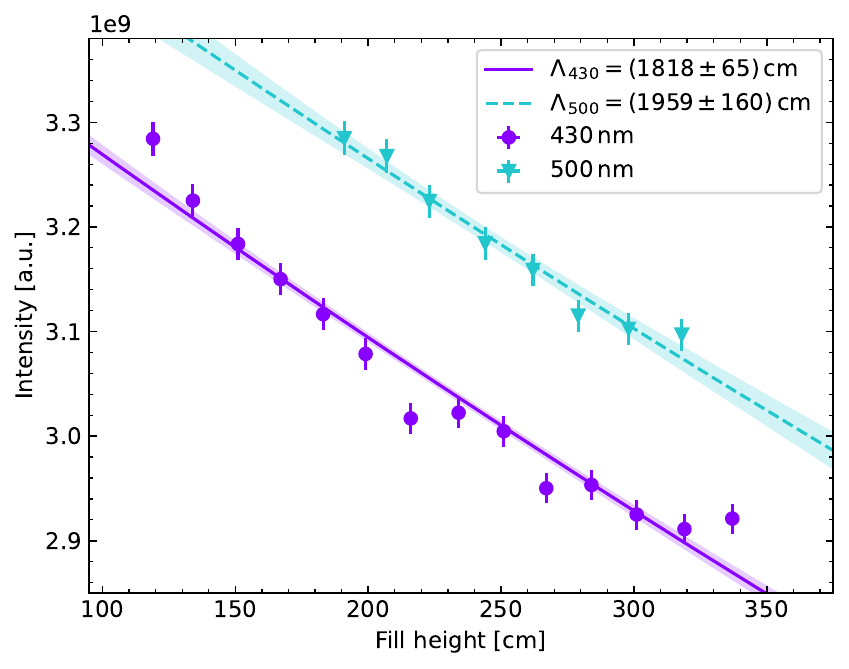}
    \caption{The measured intensity at different fill heights for Sasol LAB at \SI{430}{\nm} and \SI{500}{\nm} with the Beer-Lambert fit.}
    \label{fig:sasol_lab_430_fit}
\end{figure}

The measurements were conducted by decreasing the fill height, as it has been found that this brings higher stability. For every fill height, five intensity measurements were taken to increase the statistics. To obtain the attenuation length a fit of the Beer-Lambert-Law (Equation \ref{equ:Lambert}) is performed on the data. The uncertainties were then calculated as described in \mbox{Chapter \ref{sec:uncertainties}}.\\ Pure linear alkylbenzene (LAB), without any wavelength shifters, is used for test measurements of the PALM setup as this liquid is often used in large volume neutrino experiments, such as the upcoming~JUNO project. For this purpose, we are using unpurified LAB produced by the company Sasol. The measurements were performed at \SI{430}{\nm} and \SI{500}{\nm}. \mbox{Figure \ref{fig:sasol_lab_430_fit}} shows the Beer-Lambert fit of the measured intensities at different fill heights for LAB. The fit yields an attenuation length of $\Lambda_{430} =(1818 \, \pm \, 65)\,\text{cm}$ and $\Lambda_{500} = (1959 \, \pm \, 160)\,\text{cm}$.

\begin{figure}[htbp]
    \centering
    \includegraphics[width=0.8\textwidth]{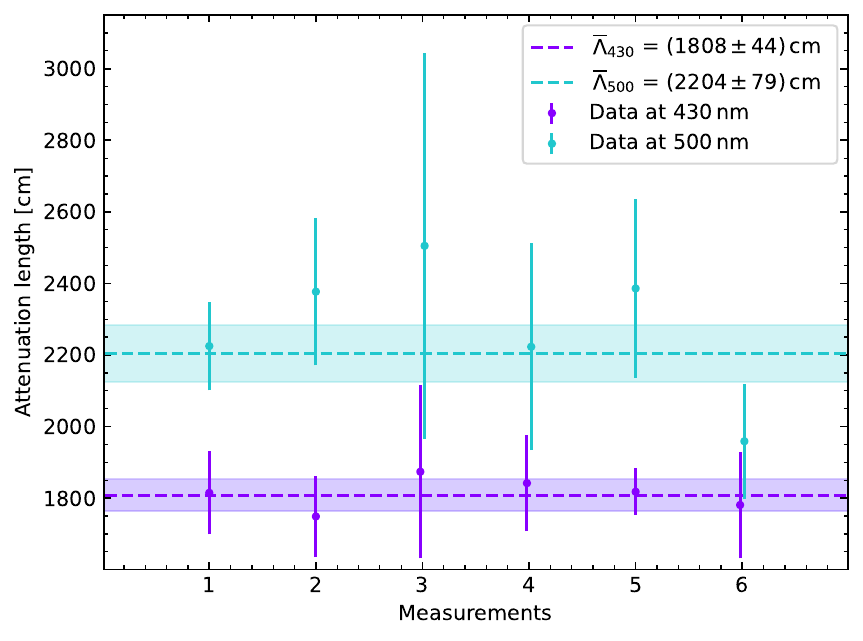}
    \caption{Overview of the attenuation length of six measurements for unpurified Sasol LAB at \SI{430}{\nm} and \SI{500}{\nm}. The dashed lines represent the weighted means, with their respective uncertainties depicted as shaded bands.}
    \label{fig:sasol_lab_overview}
\end{figure}

To verify the stability of the setup, the measurements were repeated six times under the same conditions. \mbox{Figure \ref{fig:sasol_lab_overview}} shows the results for the attenuation lengths, including the corresponding weighted mean and its associated uncertainty. The results demonstrate consistency within their uncertainties, confirming the stability of the setup. The weighted mean of the measurements at~\SI{430}{\nm} is $\overline{\Lambda_{430}} = (1808 \pm 44)\, \text{cm}$ and at \SI{500}{\nm} $\overline{\Lambda_{500}} = (2204 \pm 79)\, \text{cm}$

%\begin{table}[htbp]
%\centering
%\begin{tabular}{c|c}
%\hline
%Measurement & Attenuation Length\\ \hline 
%1 & $(1815 \, \pm \, 116)\,\text{cm}$ \\ 
%2 & $(1749 \, \pm \, 114)\,\text{cm}$ \\
%3 & $(1874 \, \pm \, 242)\,\text{cm}$ \\ 
%4 & $(1842 \, \pm \, 134)\,\text{cm}$ \\ 
%5 & $(1818 \, \pm \, 65 )\,\text{cm}$ \\
%6 & $(1781 \, \pm \, 149)\,\text{cm}$ \\ \hline
%\end{tabular}
%\caption{Results for the attenuation length of Sasol LAB at \SI{430}{\nm} for six measurements. All values are consistent within the uncertainties verifying the stability of the setup. \label{tab:sasol_lab_430_consistency}}
%\end{table}

\section{Uncertainties}
\label{sec:uncertainties}
The primary source of uncertainties was attributed to external vibrations however, the implementation of the floating gauge effectively resolved this issue. 
The total error is composed of statistical and systematic uncertainties. The value for the fill height is read from a tape measure at every measurement point. Its uncertainty is conservatively estimated to be $\Delta l_{\text{syst}} = \SI{0.5}{\cm}$.
\\ For each fill height, five intensity measurements are averaged. Their statistical uncertainty is calculated by $\Delta I_{\text{stat}} = t(n) \cdot \sigma_I$, where $\sigma_I$ denotes the standard deviation from the mean intensity and $t$ is the Student's t correction factor for $n$ measurements.
\\ To determine the fluctuation of the transmission, the tube was filled with Sasol LAB, and the intensity fluctuations were measured in three runs. They were conducted at \SI{500}{\nm} for a higher light intensity. The standard deviation of all measurements $\sigma_{\text{tot}}$ is used as a measure for the systematic uncertainty $\Delta I_{\text{syst}} = \sigma_{\text{tot}} = \SI{0.49}{\percent}$. The total uncertainty of the intensity is then obtained via the square sum rule.
\\ The error bands of the fit curve are calculated with the bootstrapping method. The relative error is proportional to $N^{-1/2}$, where $N$ is the number of samples. Since it is caused by the sampling, it can be reduced by increasing the number of samples. For this analysis, the errors are computed from~1000 samples.

\section{Discussion }
\label{sec:discussion}

The PALM spectrometer was developed to precisely measure the attenuation length of highly transparent samples across a wide range of wavelengths between about \SI{400}{\nm} and \SI{1000}{\nm}. The setup uses a halogen light source with a monochromator and the beam is focused by a Cassegrain mirror telescope. Then, the light passes through a long sample tube with adjustable fill height. The attenuation length of the sample is given by the function of the Beer-Lambert-Law. With the~PALM setup, it is feasible to measure attenuation lengths up to about \SI{100}{\m}. PALM employs optical imaging to ensure both the quality of the beam and the accuracy of the measurement throughout the measurement process.
The precision and stability of PALM were verified through a calibration process, including the gain calibration and linearity tests of the CMOS camera. Tests with an unpurified Sasol LAB sample, as detailed in Section \ref{sec:results}, successfully demonstrate the functionality of the measurement principle and the precision of the setup. PALM yields consistent and reproducible attenuation length measurements, with reliable values obtained at \SI{430}{\nm} and \SI{500}{\nm} wavelengths. The attenuation length of unpurified Sasol LAB was measured by PALM to be $\Lambda_{430} = (1808 \pm 44)\, \text{cm}$ and $\Lambda_{500} = (2204 \pm 79)\, \text{cm}$, determined from each six consistent measurements. The systematic error of the individual measurement is estimated via stability measurements. Therefore, PALM is fully prepared to accurately measure the attenuation lengths of high transparency samples.

\acknowledgments

This work benefited substantially from the support and funding by the Cluster of Excellence Universe. Further funding by the DFG Research Units FOR 5519 "Precision Neutrino Physics with~JUNO" and FOR 2319 "Determination the neutrino mass hierarchy with the~JUNO experiment" supported the project greatly. %For countless detailed and inspiring discussions, we would like to thank especially apl. Prof. Dr. Andreas Ulrich and Prof. Dr. Franz von Feilitzsch.
We would like to thank Dr. Dominikus Hellgartner for his important contributions to the later calibration of the light sensor chip of our cameras. Special thanks also go to the team at the precision engineering workshop of the Chair of Experimental Astroparticle Physics (E15) at the Technical University of Munich. Inspiring ideas and excellently constructed components have brought the project forward.

% Bibliography

%% [A] Recommended: using JHEP.bst file
%% \bibliographystyle{JHEP}
%% \bibliography{biblio.bib}

%% or
%% [B] Manual formatting (see below)
%% (i) We suggest to always provide author, title and journal data or doi:
%% in short all the informations that clearly identify a document.
%% (ii) please avoid comments such as "For a review'', "For some examples",
%% "and references therein" or move them in the text. In general, please leave only references in the bibliography and move all
%% accessory text in footnotes.
%% (iii) Also, please have only one work for each \bibitem.

%\begin{thebibliography}{99}

\bibliographystyle{JHEP}
\bibliography{main.bib}

%\bibitem{a}
%Author,
%\emph{Title},
%\emph{J. Abbrev.} {\bf vol} (year) pg.

%\bibitem{b}
%Author,
%\emph{Title},
%arxiv:1234.5678.

%\bibitem{c}
%Author,
%\emph{Title},
%Publisher (year).

%\end{thebibliography}

\end{document}